# THERMODYNAMIC PROPERTIES OF IMPROVED DEFORMED EXPONENTIAL-TYPE POTENTIAL (IDEP) FOR SOME DIATOMIC MOLECULES

Uduakobong S. Okorie[*], Akpan N. Ikot, Ephraim O. Chukwuocha

Department of Physics, Theoretical Physics Group, University of Port Harcourt, Choba, Nigeria.

## Abstract

Within the framework of non-relativistic quantum mechanics, the ro-vibrational energy spectra of the improved deformed exponential-type potential model are obtained using the Greene-Aldrich approximation scheme and an appropriate coordinate transformation. With the help of the energy spectra, analytical expressions of the vibrational partition function and other thermodynamic functions are derived by employing the Poisson summation formula. These thermodynamic functions are studied for the electronic states of hydrogen dimer, carbon monoxide, nitrogen dimer and lithium hydride diatomic molecules, as they vary with temperature and upper bound vibration quantum number.

**Corresponding Author's Email:** uduakobongokorie@aksu.edu.ng

Other Author's Email: ndemikotphysics@gmail.com, Ephraim.chukwuocha@uniport.edu.ng

## 1 Introduction

The study of statistical physics in general and quantum statistical mechanics in particular has over the years, made it possible to predict and interpret different thermodynamic properties of various systems [1]. These have contributed to the understanding of both relativistic and nonrelativistic wave equations, which contains much of the information about any given system [2]. The system of our interest is a potential energy function, which have been studied for several decades now. Example of such system is the exponential-type potential functions [3-5] and these have been examined by many authors. Recently, improved deformed four-parameter exponential-type potential energy model [6] was generated using dissociation energy and equilibrium bond length as explicit parameters for diatomic molecules. The solution of the D-dimensional Klein-Gordon equation (KGE) for multiparameter exponential-type potential has been obtained via supersymmetric quantum mechanics [7]. The improved five-parameter exponential-type potential [8] has been shown to be identical with the Tietz potential [9] for diatomic molecules. Also, the improved five-parameter exponential-type potential model have been used to model internuclear interaction potential curve for diatomic molecules using experimental molecular constants. An improved multiparameter exponential-type potential (MPETP) [10] for diatomic molecules have also been reported.

The vibrational and rotational energy levels of different potential energy models [11-14] have been established to play a key role in determining partition function of any system [15]. These energy levels can be obtained using different techniques [16-20]. Different researchers have studied the thermodynamic properties of different potential functions for various systems [21-23]. The thermodynamic properties of the improved Tietz potential was calculated for gaseous substances [24] using the Poisson summation formula [25]. Under the influence of external magnetic and electric fields, the thermodynamic properties of gallium arsenide (GaAs) double ring-shaped quantum dot have been investigated [26] via the canonical ensemble approach. A closed-form vibrational partition function and other thermodynamic functions of q-deformed Morse potential have been studied in one-dimension [27]. In addition, the thermodynamic properties of generalized Morse potential (GMP) have been derived and studied for hydrogen chloride molecule [28]. The thermodynamic functions of the Klein-Gordon equation (KGE) with the Badawi-Bessis-Bessis (BBB) diatomic molecular potential have been obtained for lithium dimer [29]. By employing the Euler

MacLaurin Formula [30, 31], the thermal properties of the Morse potential have been investigated for some diatomic molecules [32] in three dimensions.

In this work, the improved deformed exponential-type potential (IDEP) model to be employed is defined as

$$V_{IDEP}(r) = D_e \left(1 - \frac{q - e^{2\alpha(r_e - r_0)}}{q - e^{2\alpha(r - r_0)}}\right)^2 \qquad (1)$$

Here, $D_e$ is the dissociation energy, $r_e$ is the equilibrium bond length, $q$, $\alpha$ and $r_0$ are parameters that can be defined in terms of the diatomic molecular constants with the following relations respectively:

$$q = \left(1 - \frac{\alpha}{\pi c \omega_e}\sqrt{\frac{2D_e}{\mu}}\right) e^{2\alpha(r_e - r_0)} \qquad (2)$$

$$\alpha = 3\pi c \omega_e \sqrt{\frac{\mu}{2D_e}} - \frac{16\pi^3 c^2 \mu^2 r_e^3 \omega_e \alpha_e}{3h^2} - \frac{1}{2\pi r_e} \qquad (3)$$

$$r_0 = r_e - \frac{1}{9\pi c \omega_e \sqrt{\frac{\mu}{2D_e}} - \frac{16\pi^3 c^2 \mu^2 r_e^3 \omega_e \alpha_e}{h^2} - \frac{3}{2\pi r_e}} \ln\left[\frac{32\pi^3 c^2 \mu^2 r_e^4 \omega_e^2 \alpha_e + \frac{3h^2 \omega_e}{\pi}}{3h^2 + 2\pi c r_e \omega_e}\right] \qquad (4)$$

where $\omega_e$ is the equilibrium harmonic vibrational frequency, $\alpha_e$ is the vibrational rotation coupling constant, $c$ is the speed of light, $h$ is the Planck constant and $\mu$ is the reduced mass of a diatomic molecule. It is obvious that the IDEP and the improved Tietz potential functions are identical, due to the fact that replacing $2\alpha$ by $\alpha$ and $-qe^{2\alpha r_0}$ by $h$, respectively in Eq. (1) reduces the IDEP to the improved Tietz potential [33]. Hence, these potential functions can be used to produce the same fitted parameter values for different diatomic molecules. The improved Tietz potential and other related improved diatomic molecule potentials have been employed to represent the internal vibration of many diatomic and triatomic molecules, calculate the thermodynamic and thermochemical properties of these molecules [34-38].

We are motivated to evaluate the thermodynamic functions of the IDEP in terms of temperature and upper bound quantum number for some diatomic molecules, which does not exist in literatures to the best of our knowledge. The paper is organized as follows: In section 2, we determine the eigensolutions of the IDEP by employing appropriate approximation and coordinate transformation schemes. Section 3 is devoted to deriving analytical expressions for the thermal functions of the IDEP using the Poisson summation formula. Section 4 discusses the results obtained as it is applicable to hydrogen dimer, carbon monoxide, nitrogen dimer and lithium hydride diatomic molecules. The conclusion of the work is presented in section 5.

## 2  Eigensolutions of the Schrodinger Equation with IDEP

The time-independent radial Schrodinger equation is given as [39]

$$-\frac{\hbar^2}{2\mu}\frac{d^2 R_{n\ell}(r)}{dr^2} + \left[V(r) + \frac{\ell(\ell+1)\hbar^2}{2\mu r^2}\right]R(r) = E_{n\ell}R_{n\ell}(r) \qquad (5)$$

where $\mu$ is the mass, $E_{n\ell}$ is the energy spectrum of the IDEP to be determined, $\hbar$ is the reduced Planck's constant and $n$ and $\ell$ are the vibrational and rotational quantum numbers, respectively.

Substituting Eq. (1) into Eq. (5) results

$$-\frac{\hbar^2}{2\mu}\frac{d^2 R_{n\ell}(r)}{dr^2} + \left[D_e\left(1 - \frac{q - e^{2\alpha(r_e - r_0)}}{q - e^{2\alpha(r - r_0)}}\right)^2 + \frac{\ell(\ell+1)\hbar^2}{2\mu r^2}\right]R_{n\ell}(r) = E_{n\ell}R_{n\ell}(r) \qquad (6)$$

It is noted that Eq. (6) cannot be solved analytically for the case $\ell \neq 0$, due to the presence of the centrifugal term $\frac{\ell(\ell+1)}{r^2}$. As such, we employ the improved Greene - Aldrich approximation scheme to deal with the centrifugal term. This approximation scheme is given as [40]

$$\frac{1}{r^2} = 4\alpha^2\left(d_0 + \frac{e^{2\alpha r}}{\left(Q - e^{2\alpha r}\right)^2}\right), \quad d_0 = \frac{1}{12} \qquad (7)$$

Substituting Eq. (7) into Eq. (6) yields

$$-\frac{d^2 R_{n\ell}(r)}{dr^2} + \left[\left(\frac{\frac{2\mu D_e}{\hbar^2}\left(Q - e^{2\alpha r_e}\right)^2 + 4\gamma\alpha^2 e^{2\alpha r}}{\left(Q - e^{2\alpha r}\right)^2}\right) - \left(\frac{\frac{4\mu D_e}{\hbar^2}\left(Q - e^{2\alpha r_e}\right)}{\left(Q - e^{2\alpha r}\right)}\right)\right]R_{n\ell}(r)$$
$$= \frac{1}{\hbar^2}\left(2\mu E_{n\ell} - 4\gamma\hbar^2\alpha^2 d_0 - 2\mu D_e\right)R_{n\ell}(r) \qquad (8)$$

Here, we have used the following definitions: $Q = qe^{2\alpha r_0}$, $\gamma = \ell(\ell+1)$. Employing the coordinate transformation of the form $s = e^{2\alpha r}$, Eq. (8) becomes

$$s^2 \frac{d^2 R_{n\ell}(s)}{ds^2} + s \frac{dR_{n\ell}(s)}{ds} + \left[ -\varepsilon_{n\ell} + \frac{A}{(Q-s)} + \frac{B-\gamma s}{(Q-s)^2} \right] R_{n\ell}(s) = 0 \qquad (9)$$

where,

$$\varepsilon_{n\ell} = -\left[ \frac{1}{4\hbar^2 \alpha^2} \left( 2\mu E_{n\ell} - 4\gamma \hbar^2 \alpha^2 d_0 - 2\mu D_e \right) \right];$$

$$A = \frac{\mu D_e}{\hbar^2 \alpha^2} \left( Q - e^{2\alpha r_e} \right); \quad B = \frac{-\mu D_e}{2\hbar^2 \alpha^2} \left( Q - e^{2\alpha r_e} \right)^2 \qquad (10)$$

A wave function has be assumed to be of the form

$$R_{n\ell}(s) = s^\eta (Q-s)^\delta F_{n\ell}(s) \qquad (11)$$

where,

$$\eta = \pm \sqrt{\varepsilon_{n\ell} - \left( \frac{A}{Q} + \frac{B}{Q^2} \right)}; \quad \delta = \frac{1}{2}\left( 1 \pm \sqrt{1 - 4\left( \frac{B}{Q^2} - \frac{\gamma}{Q} \right)} \right) \qquad (12)$$

Substituting Eq. (11) into Eq. (9), we obtain

$$s(Q-s)\frac{d^2 F_{n\ell}(s)}{dF^2} + \left[ Q(1+2\eta) - (1+2\eta+2\delta)s \right] \frac{dF_{n\ell}(s)}{dF}$$
$$- \left[ \left( \eta + \delta - \sqrt{\varepsilon_{n\ell}} \right)\left( \eta + \delta + \sqrt{\varepsilon_{n\ell}} \right) \right] F_{n\ell}(s) = 0 \qquad (13)$$

Eq. (13) is a hypergeometric equation and its solution is the hypergeometric function given in the form [41]

$$F_{n\ell}(s) = {}_2F_1(a_1, b_1; c_1; s) = \frac{\Gamma(c_1)}{\Gamma(a_1)\Gamma(b_1)} \sum_{t=0}^{\infty} \frac{\Gamma(a_1+t)\Gamma(b_1+t)}{\Gamma(c_1+t)} \frac{s^t}{t!} \qquad (14)$$

where,

$$a_1 = \eta + \delta - \sqrt{\varepsilon_{n\ell}};$$
$$b_1 = \eta + \delta + \sqrt{\varepsilon_{n\ell}}; \qquad (15)$$
$$c_1 = Q(1+2\eta)$$

The hypergeometric function $F_{n\ell}(s)$ will become a polynomial of a certain degree when either $a_1$ or $b_1$ is equal to a negative integer $(-n)$. This leads to a finite hypergeometric function under the following quantum condition,

$$a_1 = -n, \quad n = 0, 1, 2, 3, \ldots, n_{\max} \qquad (16)$$

where,

$$n_{max} = \sqrt{\frac{\mu D_e (Q - e^{2\alpha r_e})}{\hbar^2 \alpha^2 Q}\left(1 - \frac{(Q - e^{2\alpha r_e})}{2Q}\right)} - \frac{1}{2}\left(1 \pm \sqrt{\frac{(Q+2\ell)^2}{Q^2} + \frac{2\mu D_e (Q - e^{2\alpha r_e})^2}{\hbar^2 \alpha^2 Q^2}}\right) \quad (17)$$

Substituting Eqs. (13) and (16) into Eq. (15) and carrying out simple algebra gives

$$\varepsilon_{n\ell} = \left[\frac{(n+\delta)}{2} + \frac{\chi}{2(n+\delta)}\right]^2 \quad (18)$$

where,

$$\chi = \left(\frac{A}{Q} + \frac{B}{Q^2}\right) \quad (19)$$

Substituting Eqs. (10), (12) and (18) into Eq. (19) results in the ro-vibrational energy spectra for the improved deformed exponential-type potential (IDEP) as,

$$E_{n\ell} = D_e + \frac{2\hbar^2 \alpha^2}{\mu} \frac{\ell(\ell+1)}{12}$$

$$-\frac{\hbar^2 \alpha^2}{2\mu}\left[\frac{2n+1 \pm \sqrt{\frac{(Q+2\ell)^2}{Q^2} + \frac{2\mu D_e (Q - e^{2\alpha r_e})^2}{\hbar^2 \alpha^2 Q^2}}}{4} + \frac{\frac{\mu D_e (Q - e^{2\alpha r_e})}{\hbar^2 \alpha^2 Q}\left(1 - \frac{(Q - e^{2\alpha r_e})}{2Q}\right)}{2n+1 \pm \sqrt{\frac{(Q+2\ell)^2}{Q^2} + \frac{2\mu D_e (Q - e^{2\alpha r_e})^2}{\hbar^2 \alpha^2 Q^2}}}\right]^2$$

(20)

In Eq. (20) above, + and − corresponds to $Q < 0$ and $Q > 0$, respectively.

Employing Eqs. (15) and (16), the unnormalized wave function $R_{n\ell}(r)$ can be written as

$$R_{n\ell}(r) = N_{n\ell} s^\eta (Q-s)^\delta {}_2F_1\left(-n, n+2(\eta+\delta); Q(1+2\eta); s\right), \quad s = e^{2\alpha r} \quad (21)$$

In terms of Jacobi Polynomials, we have

$$R_{n\ell}(r) = N_{n\ell} s^\eta (Q-s)^\delta P_n^{(2\eta, 2\delta-1)}(Q-2s), \quad s = e^{2\alpha r} \quad (22)$$

where $N_{n\ell}$ is the normalization constant.

The following normalization condition is employed to evaluate the normalization constant [41]

$$\int_0^\infty \left|R_{n\ell}(r)\right|^2 dr = 1 \tag{23}$$

Employing Eq. (22) into Eq. (23), we have

$$\left[R_{n\ell}(r)\right]^2 = N_{n\ell}^2 s^{2\eta} (Q-s)^{2\delta} \left[P_n^{(2\eta,2\delta-1)}(Q-2s)\right]^2 \tag{24}$$

Employing the normalization condition, we have

$$\frac{N_{n\ell}^2}{2\alpha} \int_1^0 s^{2\eta} (Q-s)^{2\delta} \left[P_n^{(2\eta,2\delta-1)}(Q-2s)\right]^2 \frac{ds}{s} = 1, \quad s = e^{2\alpha r} \tag{25}$$

Carrying out a coordinate transformation $z = Q - 2s$, Eq. (25) becomes

$$-\frac{N_{n\ell}^2}{4\alpha} \int_{(Q-2)}^{Q} \left(\frac{Q-z}{2}\right)^{2\eta-1} \left(\frac{Q+z}{2}\right)^{2\delta} \left[P_n^{(2\eta,2\delta-1)}(z)\right]^2 dz = 1 \tag{26}$$

We now take the parameter $Q = 1$ for purposes of simplicity to have

$$-\frac{N_{n\ell}^2}{4\alpha} \int_{-1}^{1} \left(\frac{1-z}{2}\right)^{2\eta-1} \left(\frac{1+z}{2}\right)^{2\delta} \left[P_n^{(2\eta,2\delta-1)}(z)\right]^2 dz = 1 \tag{27}$$

Using the standard integral [41],

$$\int_{-1}^{1} \left(\frac{1-w}{2}\right)^x \left(\frac{1+w}{2}\right)^y \left[P_n^{(x,y-1)}(w)\right]^2 dw = \frac{2^{x+y+1}\Gamma(x+n+1)\Gamma(y+n+1)}{n!\,\Gamma(x+y+n+1)\Gamma(x+y+2n+1)} \tag{28}$$

Comparing Eqs. (27) and (28), where $x = 2\eta - 1$ $and$ $y = 2\delta$, the normalization constant is obtained as

$$N_{n\ell} = \sqrt{\frac{(-1)4\alpha(n!)\Gamma(n+2\delta+2\eta)\Gamma(2n+2\delta+2\eta)}{2^{(2\delta+2\eta)}\Gamma(n+2\eta)\Gamma(1+n+2\delta)}} \tag{29}$$

## 3    Thermal Functions of the IDEP

The vibrational partition function is the beginning point to determine any thermal function of a system [42]. The bound state contributions to the vibrational partition function of any system at a given temperature T is defined as

$$Z(\beta, \lambda) = \sum_{n=0}^{\lambda} e^{-\beta E_{n\ell}}, \quad \beta = (k_B T)^{-1} \quad (30)$$

where $k_B$ is the Boltzmann's constant, $\lambda$ is the upper bound quantum number, $E_{n\ell}$ is the ro-vibrational energy eigenvalues of the IDEP. First, we set $\ell = 0$ and simplify Eq. (20) to have

$$E_n = D_e - \frac{\hbar^2 \alpha^2}{2\mu} \left[ \frac{(n+\varsigma)}{2} + \frac{\chi}{2(n+\varsigma)} \right]^2 \quad (31)$$

where

$$\chi = \frac{\mu D_e (Q - e^{2\alpha r_e})}{2\hbar^2 \alpha^2 Q} \left( 1 - \frac{(Q - e^{2\alpha r_e})}{2Q} \right) \quad (32)$$

$$\varsigma = \frac{1}{2} \left( 1 \pm \sqrt{1 + \frac{2\mu D_e (Q - e^{2\alpha r_e})^2}{\hbar^2 \alpha^2 Q^2}} \right) \quad (33)$$

The vibrational partition function can be calculated by summing all possible vibrational energy levels of a given system directly. Substituting Eq. (31) into Eq. (30) results in the following

$$Z(\beta, \lambda) = \sum_{n=0}^{\lambda} \exp \left\{ -\beta \left( D_e - \frac{\hbar^2 \alpha^2}{2\mu} \left( \frac{(n+\varsigma)}{2} + \frac{\chi}{2(n+\varsigma)} \right)^2 \right) \right\} \quad (34)$$

In evaluating Eq. (34), we employ the Poisson summation formula for lower order approximation as

$$\sum_{n=0}^{n_{\max}} f(x) = \frac{1}{2} [f(0) - f(n_{\max}+1)] + \int_0^{n_{\max}+1} f(y) dy \quad (35)$$

By substituting Eq. (34) into Eq. (35), one can obtain the following functions:

$$f(0) = \exp\{-\beta(D_e - H G_1^2)\}; \quad f(\lambda+1) = \exp\{-\beta(D_e - H G_2^2)\} \quad (36)$$

where

$$H = \frac{\hbar^2 \alpha^2}{2\mu}; \quad G_1 = \frac{\varsigma}{2} + \frac{\chi}{2\varsigma}; \quad G_1 = \frac{(\lambda + 1 + \varsigma)}{2} + \frac{\chi}{2(\lambda + 1 + \varsigma)} \tag{37}$$

Also,

$$\int_0^{\lambda+1} f(x)dx = \int_0^{\lambda+1} \exp\left\{-\beta\left(D_e - \frac{\hbar^2 \alpha^2}{2\mu}\left(\frac{(\varpi + \varsigma)}{2} + \frac{\chi}{2(\varpi + \varsigma)}\right)^2\right)\right\} d\varpi, \quad \varpi = \lambda + 1 \tag{38}$$

Evaluating the integral part of the right-hand side of Eq. (38), we have

$$\int_0^{\lambda+1} f(x)dx = \int_{G_1}^{G_2} \exp\left\{-\beta(D_e - H\rho^2)\right\} \cdot \left(1 + \frac{\rho}{\sqrt{\rho^2 - \chi}}\right) d\rho \tag{39}$$

Here, we have defined the parameter

$$\rho = \frac{(\varpi + \varsigma)}{2} + \frac{\chi}{2(\varpi + \varsigma)} \tag{40}$$

Employing the Mathematica software to evaluate Eq. (39) and combining with Eq. (36), the vibrational partition function of IDEP is obtained as

$$Z(\beta, \lambda) = \frac{1}{2} e^{-\beta D_e} \left\{ \begin{array}{l} e^{\beta H G_1^2} - e^{\beta H G_2^2} \\ + \sqrt{\frac{\pi}{H\beta}} \left[ \begin{array}{l} -Erfi\left(G_1 \sqrt{H\beta}\right) + Erfi\left(G_2 \sqrt{H\beta}\right) \\ + e^{H\chi\beta} \left(-Erfi\left(\sqrt{H\beta}\sqrt{G_1^2 - \chi}\right) + Erfi\left(\sqrt{H\beta}\sqrt{G_2^2 - \chi}\right)\right) \end{array} \right] \end{array} \right\}$$

(41)

where the imaginary error function is defined as

$$Erfi(u) = \frac{Erf(iu)}{i} = \frac{2}{\sqrt{\pi}} \int_0^u e^{\tau^2} d\tau \tag{42}$$

With the help of the vibrational partition function of Eq. (41), other thermodynamic functions of IDEP can be obtained using the following expressions:

➤ Vibrational Internal energy

$$U(\beta, \lambda) = -\frac{\partial \ln Z(\beta, \lambda)}{\partial \beta} \tag{43}$$

➤ Vibrational free energy

$$F(\beta, \lambda) = -\frac{1}{\beta} \ln Z(\beta, \lambda) \tag{44}$$

➢ Vibrational entropy

$$S(\beta, \lambda) = k_B \ln Z(\beta, \lambda) - k_B \beta \frac{\partial \ln Z(\beta, \lambda)}{\partial \beta} \quad (45)$$

➢ Vibrational specific heat capacity

$$C(\beta, \lambda) = k_B \beta^2 \frac{\partial^2}{\partial \beta^2} \ln Z(\beta, \lambda) \quad (46)$$

## 4 Results and Discussion

In this paper, the electronic states of hydrogen dimer, carbon monoxide, nitrogen dimer and lithium hydride are considered using the ro-vibrational energy spectra of Eq. (20). The experimental values of the selected diatomic molecules are given in Table 1 [43, 44]. The values of the molecular parameters $q$, $\alpha$ and $r_0$ for the selected diatomic molecules are also obtained using Eqs. (2) – (4), respectively, as tabulated in Table 2. It is worthy to mention here that the following conversion factors have been employed throughout the computations:

$1\, cm^{-1} = 1.239841875 \times 10^{-4}\, \text{eV}; \hbar = 1973.296\, \frac{\text{eV}\, \text{Å}}{c}; 1\, amu = 931.494028\, \frac{\text{MeV}}{c^2}$. By

employing the experimental data as our input, the vibrational partition function for the selected diatomic molecules are plotted with temperature and upper bound vibration quantum number, using Eq. (41). As shown in Fig. 1, the vibrational partition function curves increase as the temperature increases for the selected diatomic molecules. The vibrational partition function curves of $H_2$ and $LiH$ are seen to increase sharply at a temperature less than $100,000\, K$. As the temperature increases beyond $100,000\, K$, the vibrational partition function curves of $H_2$ and $LiH$ remains uniquely constant. The vibrational partition function curves of $CO$ and $N_2$ is seen to increase monotonically as the temperature increases from $100,000\, K$. Beyond a temperature of $400,000\, K$, the increase in the vibrational partition function curves of $CO$ and $N_2$ remain constant. With the help of the vibrational partition function expression of Eq. (41), other thermodynamic functions variation with temperature and upper bound vibration quantum number have been evaluated using Eqs. (43) – (46). Fig. 2 displays the variation of vibrational free energy of IDEP with temperature for the selected diatomic molecules. The vibrational free energy curves for the selected diatomic molecules decreases directly as the temperature increases. In Fig. 3, a monotonous decreasing behaviour of the vibrational mean energy is observed as the temperature increases. Beyond the temperature of $200,000\ K$, the graphs for each diatomic molecule reaches almost zero asymptotically. In Fig 4, It is seen that the vibrational entropy for the selected diatomic molecules first decrease as the temperature increases from origin to $200,000\ K$. As the temperature exceeds $200,000\ K$, the vibrational entropy curves remain constant at unique values for the different diatomic molecules. Fig. 5 shows a monotonous increase in the vibrational specific heat capacity as the temperature increases for the selected diatomic molecules. Beyond a temperature of $500,000\ K$, the vibrational specific heat capacity curves for the selected diatomic molecules clusters and tends to zero.

The variations of the thermodynamic functions of IDEP with upper bound vibration quantum number are also presented in Figs. 6 -10. In Fig. 6, as the upper bound quantum number increases, the vibrational partition function curves for the selected diatomic molecules increase slowly. When the upper bound quantum number increases beyond a specific number for each diatomic molecule, the increase in the vibrational partition function curve for each diatomic molecule becomes very sharp. The sharp increase in the vibrational partition function for $H_2$ dimer is observed when the upper bound quantum number is about $50$. At an upper bound quantum number of $90$, the sharp increase in the vibrational partition function for $N_2$ dimer occurs. The sharp increase in the vibrational partition function for $CO$ and $LiH$ dimers occur when the upper bound vibration quantum number increases beyond $100$. In Fig. 7, the vibrational free energy curves for the selected diatomic molecules decrease gradually as the upper bound vibration quantum number increases. The vibrational free energy curve for $H_2$ dimer increases first gradually as the upper bound quantum number increases. Beyond the upper bound quantum number of $50$, the vibrational free energy for $H_2$ dimer decreases faster. Fig. 8 shows a sharp increase in the vibrational mean energy at an upper bound vibration quantum number of about $5$. As the upper bound quantum number increases, the vibrational mean energy curves of the selected diatomic molecules increase monotonously. In Fig. 9, the variation of the vibrational entropy with upper bound vibration quantum number for $H_2$, $CO$, $N_2$ and $LiH$ diatomic molecules is shown. As the upper bound vibration quantum number increases, the vibrational entropy is seen to increase in a gradual form for the selected diatomic molecules. We also observe a sharper increase in vibrational entropy for $H_2$ dimer as the upper bound quantum number increases. The graph of vibrational specific heat capacity of IDEP with upper bound vibration quantum number for $H_2$, $CO$, $N_2$ and $LiH$ diatomic molecules is shown in Fig. 10. At a zero upper bound vibration quantum number, the vibrational specific heat capacity of the selected diatomic molecules increases sharply to about $1\times10^{-10}\, cm^{-1}\, K^{-1}$. The increase in the upper bound vibration quantum number causes a unique gradual increase in the vibrational specific heat capacity for each of the selected diatomic molecule. The increase in the vibrational specific heat capacity for $H_2$ dimer is seen to be more sharper, as compared to $CO$, $N_2$ and $LiH$ diatomic molecules.

## 5    Conclusion

In this work, the Schrodinger equation has been solved with the improved deformed exponential-type potential (IDEP) model, by applying an appropriate approximation and a coordinate transformation schemes. The analytical expressions of the ro-vibrational energy spectra and the corresponding normalized energy eigenfunction have been obtained. Using the ro-vibrational energy spectra of the IDEP, the vibrational partition function and other thermodynamic functions like vibrational free energy, vibrational mean energy, vibrational entropy and vibrational specific heat capacity have been deduced using the Poisson summation formula. With these results, the thermodynamic properties of $H_2$, $CO$, $N_2$ and $LiH$ diatomic molecules have been studied graphically and discussed extensively, as it varies with temperature and upper bound vibration quantum number. It is of note that the IDEP can be reduced to the improved Tietz potential model.

Table 1 Spectroscopic Parameters for the selected diatomic molecules [43, 44].

| Molecules | $D_e (cm^{-1})$ | $r_e (\text{Å})$ | $\mu (amu)$ |
|---|---|---|---|
| $H_2(X^1\Sigma^+)$ | 38297.00 | 0.7416 | 0.5039 |
| $CO(X^1\Sigma^+)$ | 90529.00 | 1.1283 | 6.8562 |
| $N_2(X^1\Sigma_g^+)$ | 79885.00 | 1.0970 | 7.0034 |
| $LiH(X^1\Sigma^+)$ | 20287.70 | 1.5955 | 0.8801 |

Table 2 Calculated Values of the IDEP parameters for different diatomic molecules

| States of diatomic molecules | $(\alpha \times 10^7) cm^{-1}$ | $(r_0 \times 10^{-8}) cm$ | $q$ |
|---|---|---|---|
| $H_2(X^1\Sigma^+)$ | 25.73680374 | 0.7416001485 | $-0.3236073943$ |
| $N_2(X^1\Sigma_g^+)$ | 36.42844874 | 1.097000113 | $-0.3543700921$ |
| $LiH(X^1\Sigma^+)$ | 15.02728120 | 1.595500403 | $-0.3326882575$ |
| $CO(X^1\Sigma^+)$ | 30.69381001 | 1.128300118 | $-0.6544806294$ |

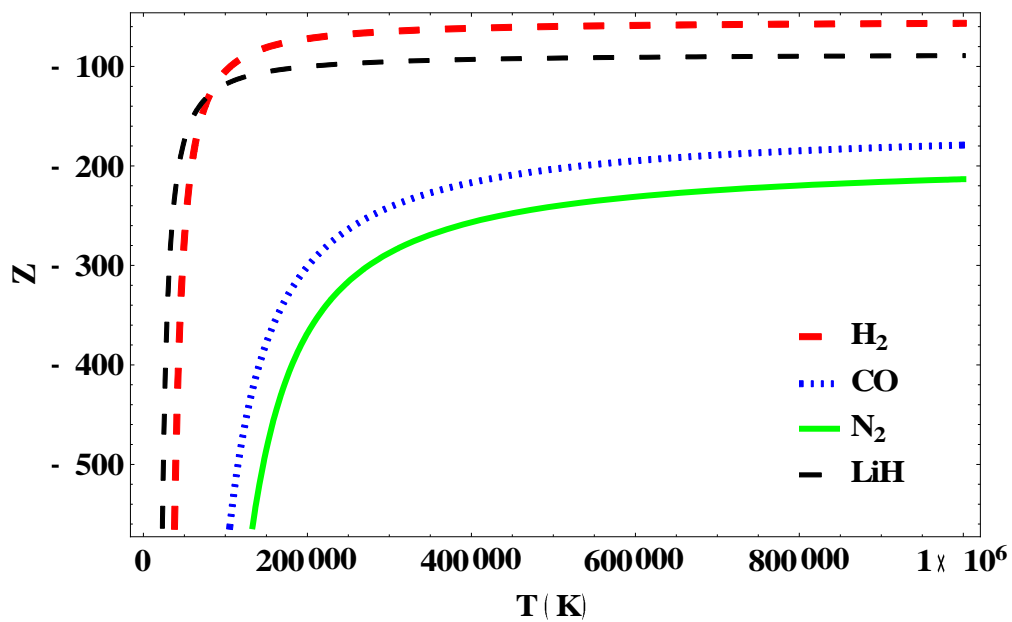

Fig. 1 Vibrational partition function versus temperature of IDEP for $H_2$, CO, $N_2$ and LiH diatomic molecules.

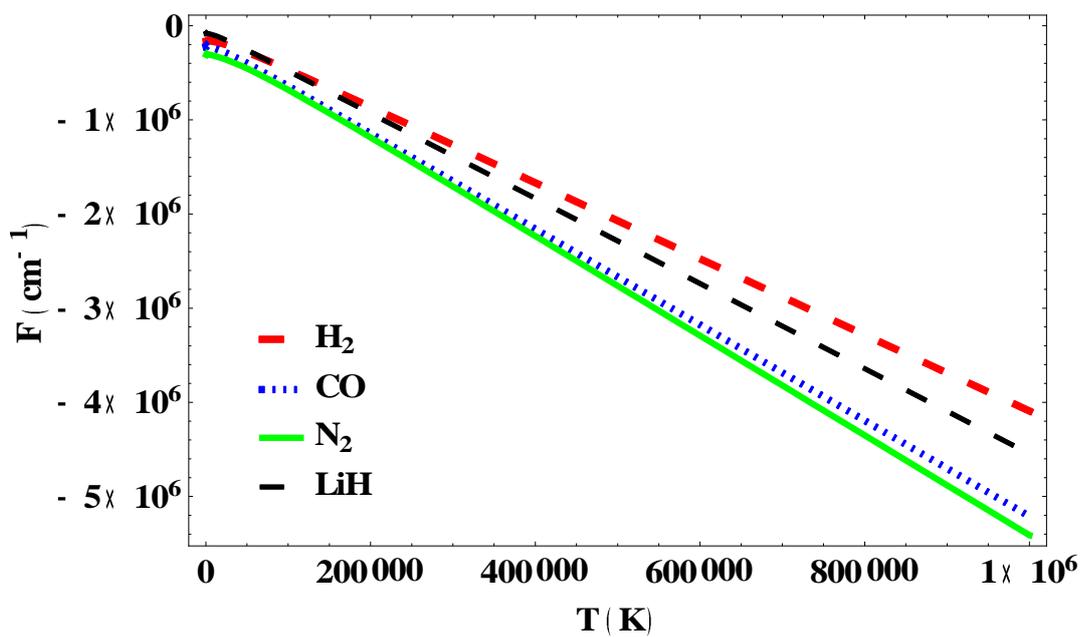

Fig. 2 Vibrational free energy versus temperature of IDEP for $H_2$, CO, $N_2$ and LiH diatomic molecules.

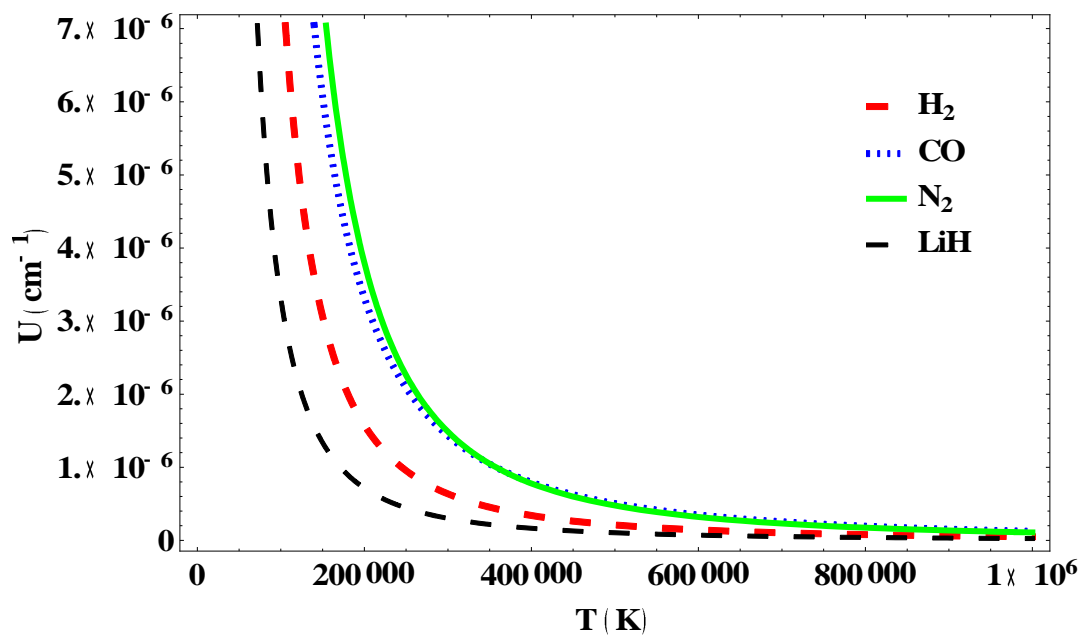

Fig. 3 Vibrational mean energy versus temperature of IDEP for $H_2$, CO, $N_2$ and LiH diatomic molecules.

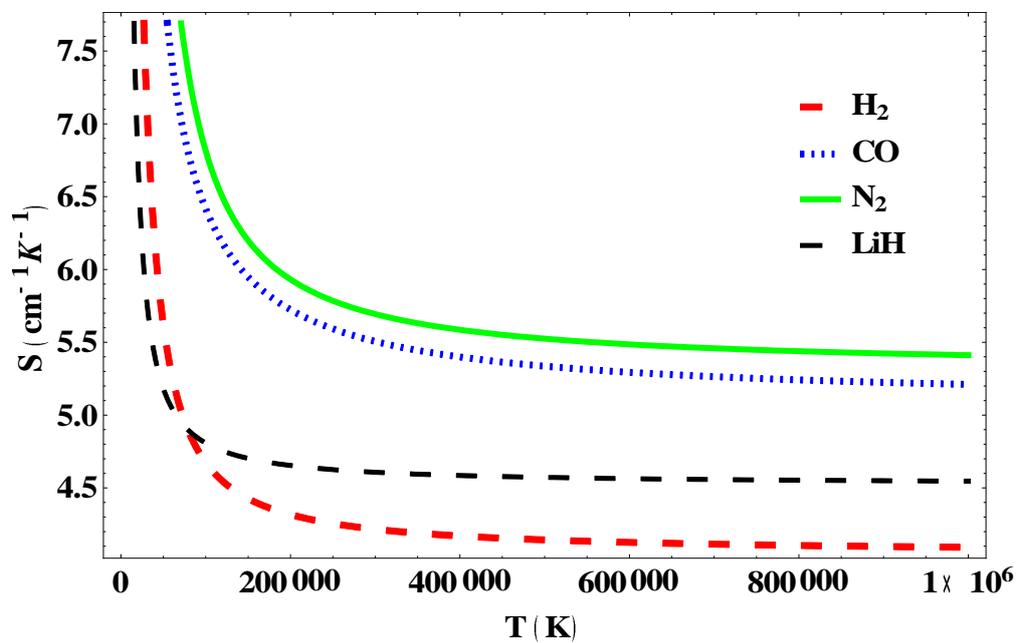

Fig. 4 Vibrational entropy versus temperature of IDEP for $H_2$, CO, $N_2$ and LiH diatomic molecules.

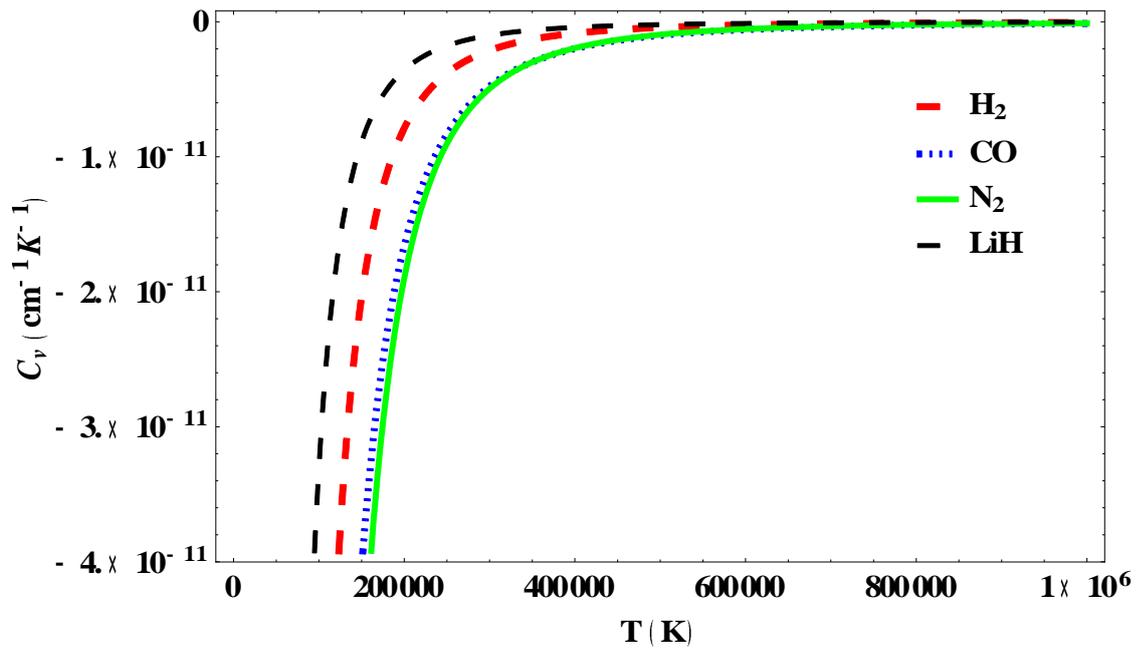

Fig. 5 Vibrational specific heat capacity versus temperature of IDEP for H2, CO, N2 and LiH diatomic molecules.

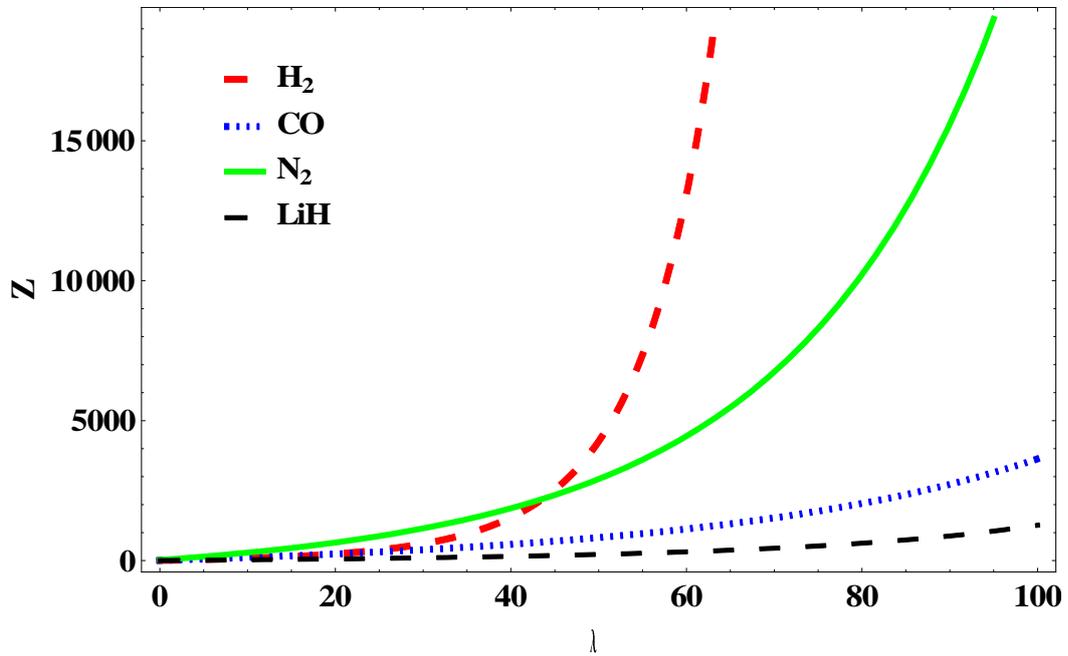

Fig. 6 Vibrational partition function of IDEP versus upper bound vibration quantum number for H2, CO, N2 and LiH diatomic molecules.

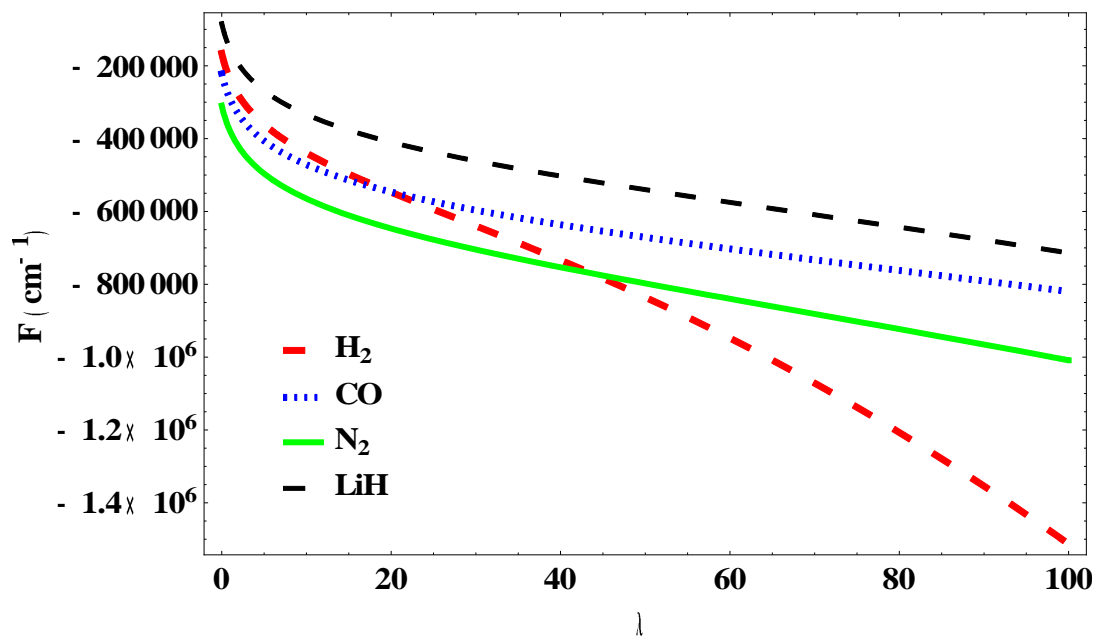

Fig. 7 Vibrational free energy of IDEP versus upper bound vibration quantum number for $H_2$, CO, $N_2$ and LiH diatomic molecules.

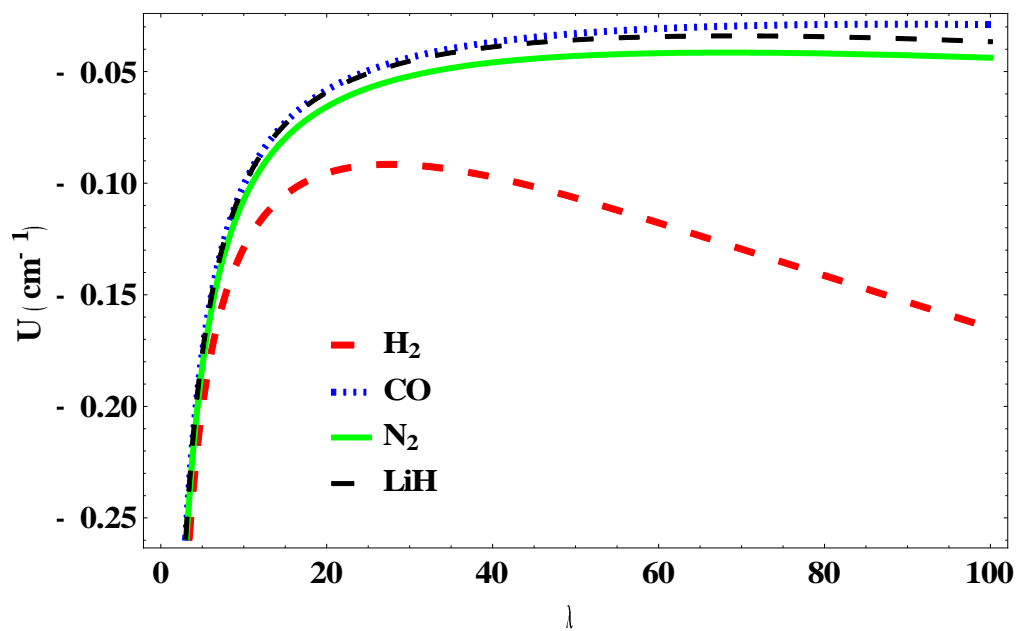

Fig. 8 Vibrational mean energy of IDEP versus upper bound vibration quantum number for $H_2$, CO, $N_2$ and LiH diatomic molecules.

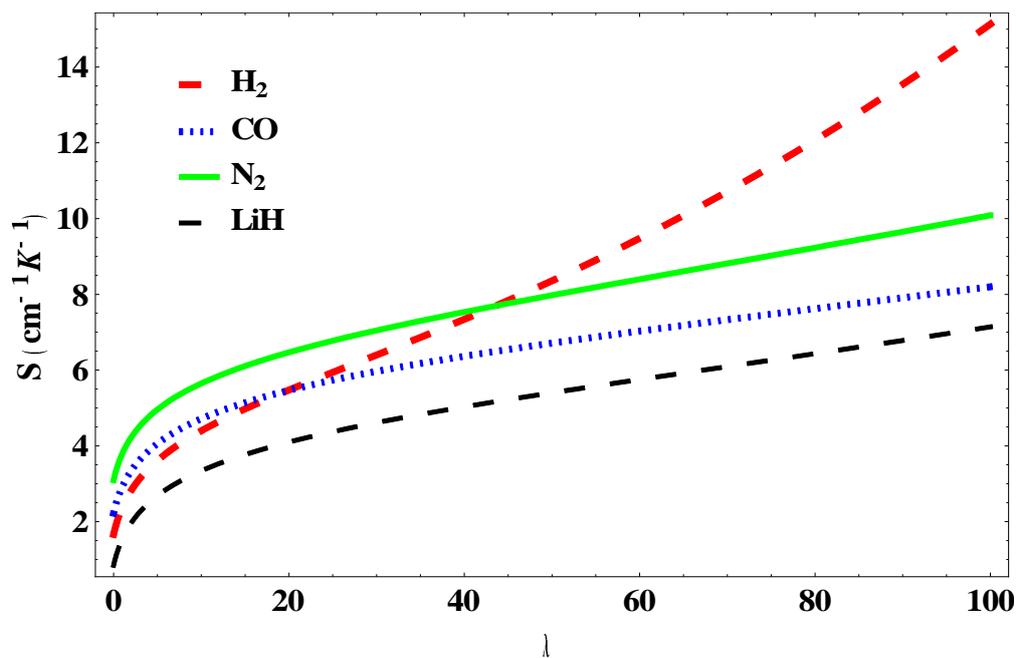

Fig. 9 Vibrational entropy of IDEP versus upper bound vibration quantum number for $H_2$, CO, $N_2$ and LiH diatomic molecules.

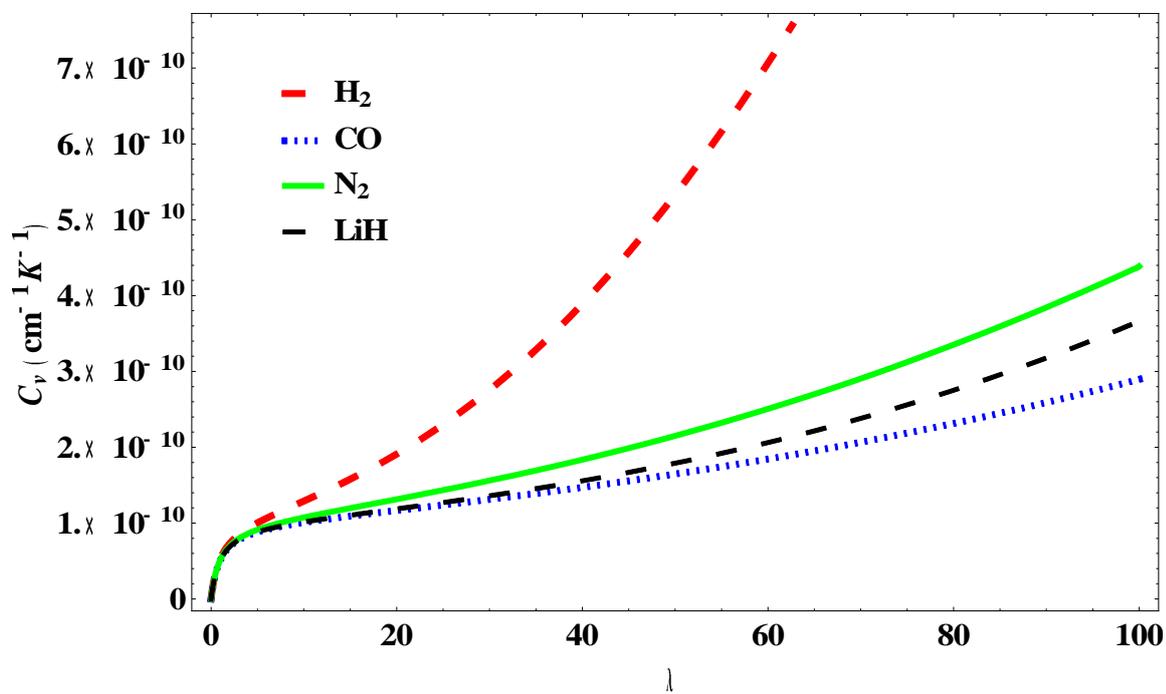

Fig. 10 Vibrational specific heat capacity of IDEP versus upper bound vibration quantum number for $H_2$, CO, $N_2$ and LiH diatomic molecules.